# Search of the Earth Limb Fermi Data and Non-Galactic Center Region Fermi Data for Signs of Narrow Lines

E. Bloom, E. Charles, E. Izaguirre, A. Snyder
*KIPAC-SLAC, Stanford University, 2572 Sand Hill Road, Menlo Park, CA, 94025 USA*

A. Albert, B. Winer, Z. Yang
*The Ohio State University, 1739 N High St, Columbus, OH 43210 USA*

R. Essig
*Stony Brook University, Stony Brook, NY 11794 USA*

**On Behalf of the Fermi-LAT Collaboration**

Since the spring of 2012 there have been many papers published using Fermi LAT public data that claim evidence for narrow spectral lines coming from the region of the Galactic center. This study uses non-Galactic center Fermi-LAT data from survey mode observations, and Earth limb Fermi data to test the dark matter interpretation of this feature and better understand its origins.

## 1. INTRODUCTION

A putative gamma ray line was reported at 130 GeV by C. Weniger [1] in the region of the Galactic center (GC) using over three years of public Pass7 Fermi-LAT data [2]. The fractional residual (or signal to noise—S/N) of this line feature is given by,

$$S/N = s^2/n_s \quad (1)$$

where **s** is the number of standard errors of the line feature above background and $n_s$ is the number of line feature events, both as determined from the fit. Typical values obtained by Weniger [1] for the 130 GeV feature give S/N ~ 0.4. In Weniger's work he selected regions of interest on the sky (ROI) to study based on estimating signal to noise using a combination of data to estimate the noise and dark matter models of the Galaxy to estimate the signal (e.g. the NFW model).

The Fermi –LAT Collaboration (LAT) has published limits based on a line search with two years of Pass 6 data [3] The ROI examined did not correspond to those determined by Weniger in his work; however, the limits are obtained in the context of the same dark matter models of the Galaxy used by Weniger. Thus LAT limits on $<\sigma^*v>[\chi\chi \to \gamma\gamma]$ can be compared to the line signal strengths of reference [1]. Table 1 shows such a comparison that indicates a mild tension between LAT limits and the Weniger line strengths.

Table 1: Comparison of LAT limits [3] and Weniger line signal strengths [1] at 130 GeV.

| DM Model | Line Strength from [1]. | LAT limits from [3] |
|---|---|---|
| NFW | (2.27±0.52 +0.32 –0.51) x$10^{-27}$ cm$^3$s$^{-1}$ | 1.4x$10^{-27}$ cm$^3$s$^{-1}$ |
| Einasto | (1.27 0.32 +0.18 –0.28) x$10^{-27}$ cm$^3$s$^{-1}$ | 1.0x$10^{-27}$ cm$^3$s$^{-1}$ |

The LAT has been working on a line search using Pass 7 data that has been recalibrated. The plenary talk of Andrea Albert at this conference reports the latest results at that time on the LAT analysis [4]. The parallel session talk of Eric Charles discusses more details of LAT systematics expected in gamma ray line searches [5]. It should be noted that the recalibration shifted the GC line feature from 130 GeV to 135 GeV [4].

## 2. SYSTEMATICS OF THE LINE ANALYSIS

### 2.1. Pass 6 Analysis

In ref [3] limits were given on narrow lines using two years of P6V3 public LAT data. That paper describes in detail a significant line seen at 6.5 GeV that was due to a systematic effect originating from the so called "PSF cut" defined in that paper. Removing the PSF cut largely removed the line, resulting in limits and no significant residuals. Two control data sets, which presumably contain little or no contribution from dark matter, were examined to show that the observed line structure seen was a systematic: the Earth limb (called the "Albedo data set" in ref [3]) and inverse ROI data sets. In the current LAT Pass 7 work, which uses roughly twice the amount of data compared to ref [3] and a new data reconstruction and calibration —P7V6 reprocessed, we again apply the lessons of ref [3] and carefully examine the Earth limb and the same inverse ROI used in [3].

### 2.2. Earth Limb Data Set

The Earth limb data set, composed of photons produced by cosmic rays scattering in the Earth's atmosphere, is obtained by changing the orbit of the LAT to observe the photons produced by cosmic rays hitting the Earth's atmosphere tangential to the surface of the Earth. Figure 1 shows a schematic of the geometry for limb gamma-ray production.

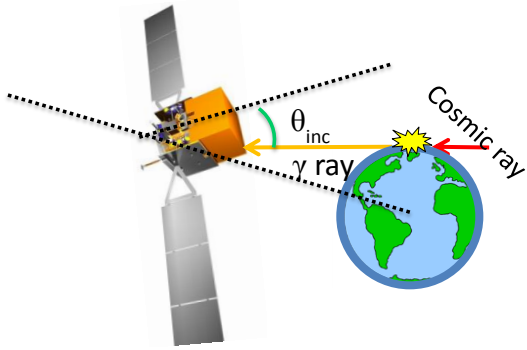

Figure 1: Schematic of Earth Limb gamma ray production by cosmic rays from the Earth's atmosphere. $\theta_{inc}$ is the theta angle of the gamma ray to the z-axis of the Large Area Telescope, $LAT\_\theta$.

For energies above 10 GeV these photons form a bright very narrow ring close to the surface of the Earth (the ring becomes ever broader for photon energies below 10 GeV). The energy spectrum of the produced photons is a power law with index ~ -2.75 [6]. The Earth limb data set used here includes P7V6 reprocessed data from August 8, 2008 thru September 2012. The cuts used to isolate the Earth limb photons in the LAT are primarily $111°$ < (Earth zenith angle) < $113°$. We use the P7V6 reprocessed CLEAN class [7] , [8]. To separate Earth limb events from normal astronomical observations an additional cut on the LAT rocking angle, LRA, is made, $|LRA| > 52°$. By making this cut, the entire Earth limb data sample is selected, but with a ratio of observing time to the entire 4 year data set of $3 \times 10^{-4}$ since the limb is so bright.

In looking for systematic effects in the limb data that might appear in the Galactic Center (GC) data, one needs to establish that the distribution of the limb photon incidence angles - $LAT\_\theta$ - on the LAT is similar to photons originating from the GC. Figure 2 shows two plots that give information about the $LAT\_\theta$ distribution. Figure 2-upper shows the observing time versus $cos(LAT\_\theta)$, as calculated from the known LAT orbit during the observations of the Earth limb and the GC (the limb curve needs to be multiplied by over 400 to be reasonably on the same scale). Figure 2-lower shows the actual distribution of gamma ray data from the GC and the limb for gamma energy in the range 75 – 200 GeV. The data from the GC and the limb track well except for $cos(LAT\_\theta) >~ 0.9$. Figure 2-upper shows a similar behavior. The differences between the GC and limb $LAT\_\theta$ dependence may impact the strength of a line feature in each data set at 135 GeV.

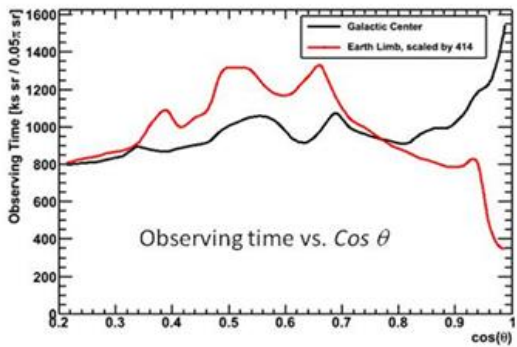
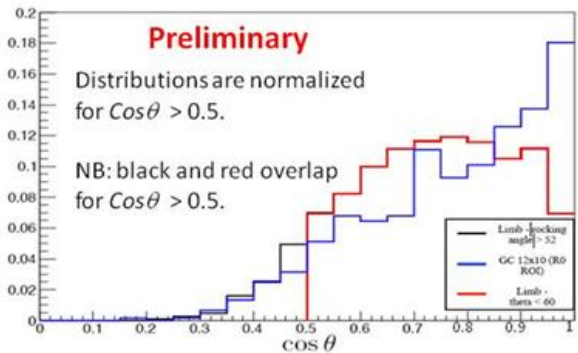

Figure 2: Plots showing the observing time versus $cos(LAT\_\theta)$, scaled by ~ 400 for the Earth limb-2 upper, and data versus $cos(LAT\_\theta)$-2 lower. The plots on the lower right are normalized for $cos(LAT\_\theta) > 0.5$ and are shown for the GC data -black, limb for $|LRA| > 52°$-blue, and additionally the limb data for $cos(LAT\_\theta) > 0.5$ – red.

Figure 3 shows an unbinned 2-D likelihood fit, done as described in [4] to a line feature plus a simple power-law background to the limb data. 2-D refers to our energy dispersion model that is a function of both energy and $P_E$, where $P_E$ is an event-by-event quantification of the accuracy of the energy reconstruction. The energy of the line feature is taken as 135 GeV as determined from the GC fit shown in [4].

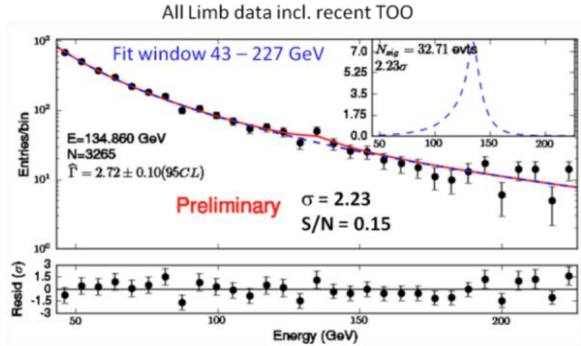

Figure 3: Unbinned 2-D likelihood fit to a line feature plus a simple power law background to the Earth limb data. The analysis uses P7V6 reprocessed_CLEAN class.

In this case the fit range is taken as 43-227 GeV from studies to determine the optimum fit range using Monte Carlo [4]. A larger fit range results in a smaller error on the background power-law, assuming that the background model is a good representation of the data, which it is for the Earth limb. The p value for the fit is 0.98, which indicates a very good fit of the model to the data. In figure 3 the S/N (equation 1) is 0.15 as compared to 0.7 obtained in [4] for the putative line feature in a 4x4 degree ROI about the GC.

Figure 4 shows an unbinned 2-D likelihood fit to the same limb data as figure 3, but cut at LAT_$\theta$ < 60°. In addition, the fit is made over the energy range of 75 GeV to 200 GeV. The significance of the line feature is enhanced in this fit to 3.35$\sigma$, and the *S/N* is somewhat larger at 0.22. The residuals to the fit indicate a somewhat narrower line feature than the LAT resolution as is the case for the GC signal discussed in [4].

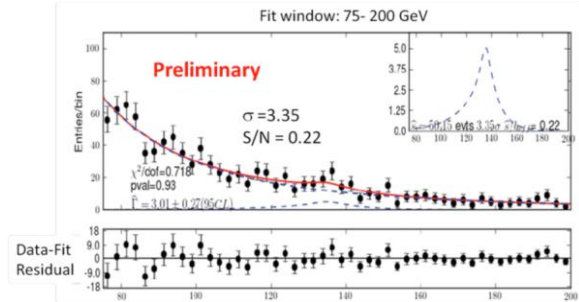

Figure 4: Same as figure 3, but with c$os(LAT\_\theta)$>0.5, and energy fit range window 75-200 GeV.

Compared to the fits in figure 3, $\Delta\sigma = 0.55$ comes about from the change in the fit window from 43- 227 GeV to 75-200 GeV. An additional $\Delta\sigma=0.59$ arises from the LAT_$\theta$ cut. It should be noted that possible "shine through" from the sky behind the limb is not an issue since the ratio of observing time for (Earth limb)/GC = $2.8\times10^{-4}$, which implies << 1 event expected from shine through.

The Earth limb evidence for a line feature at 135 GeV strongly suggests that a major part of the GC line feature is systematic in origin [5]. However, a detailed examination of the inverse ROI data set does not straight forwardly support this conclusion as little of significance is observed at 135 GeV for this data set.

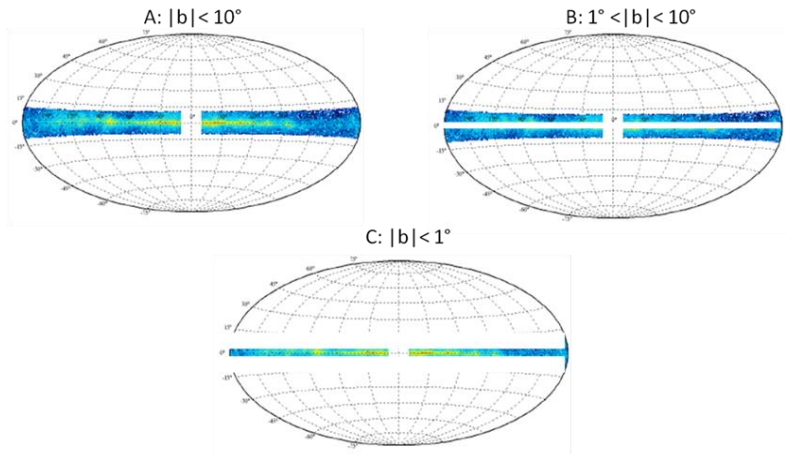

Figure 5: Definition of inverse ROI used as control data sets. A: is the Galactic plane with |b|<10° excluding ± 10° in l about the GC. B: the same as A: but excluding the stripe |b|<1°. C: is just the stripe |b|<1° excluding ± 10° in l about the GC.

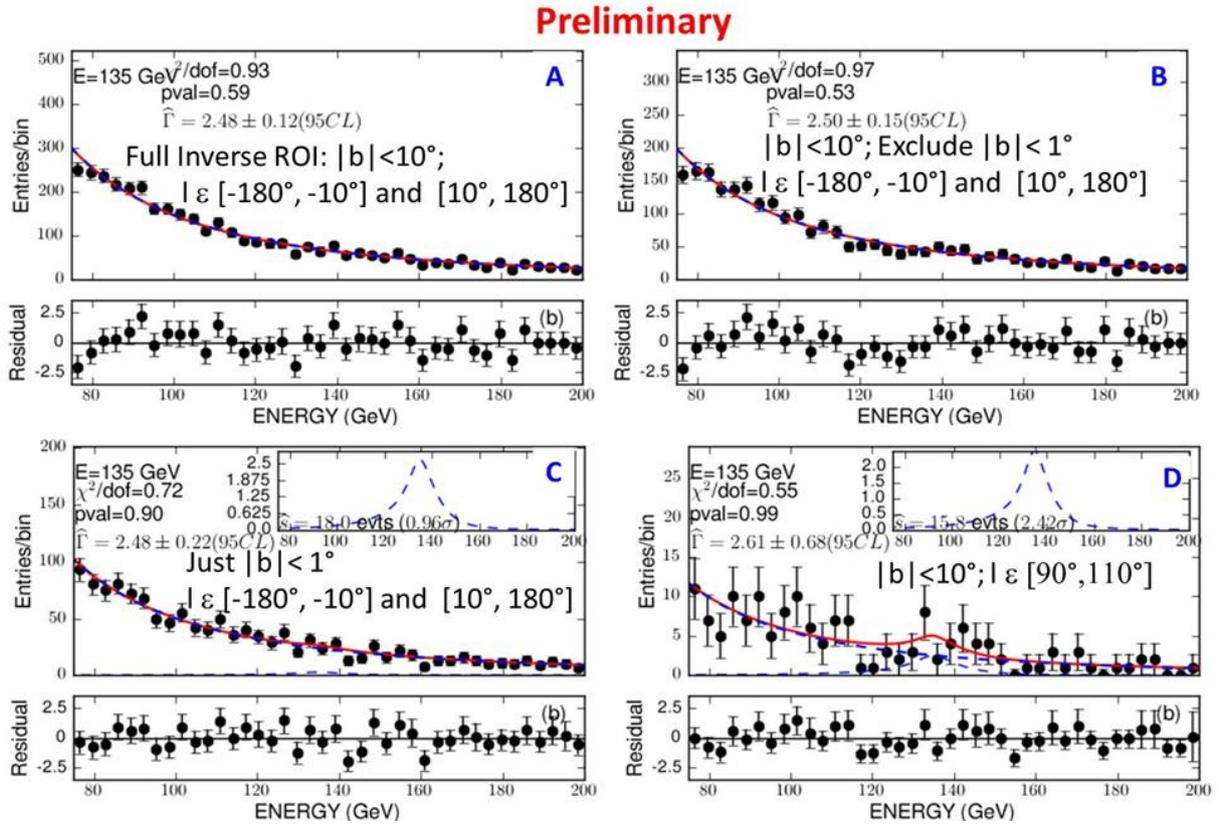

Figure 6: Overview of results fitting to the data in the control data sets of figure 6. Parts A, B, and C are corresponding to the entire region. Part D shows the fit result with the largest line feature significance of 2.4σ.

## 2.3. Inverse ROI Data Set

Figure 5 shows the definition of the inverse ROIs used in this analysis. In the figure region A: shows the full inverse ROI as defined in [2] . In addition to A: we also examined region B and C. The data set was P7V6 reprocessed CLEAN class and using the other standard analysis cuts described in [4] . First, an energy spectrum resulting from all of the data in A, in B, and in C was fit to a single power law plus a line feature fixed at 135 GeV. Figure 6 shows the results of these three fits. Parts A, B and C of figure 6 show no significant line feature at 135 GeV. In addition to integrating all the counts in each of these three regions, we also did a scan in Galactic longitude, l, for each of the regions. Each region was split into bins of 20 degrees in l, and fitted to a single power and a line feature at 135 GeV. Figure 6 D shows the fit result with the largest line feature significance of 2.4σ. Fits were made to a total of 54 20° bins over the 3 regions in l, and the result in figure 6 D had the largest significance for a line feature at 135 GeV. Thus we find no strong indication for a 135 GeV line feature in these inverse ROI control data sets.

Could there be a kinematic reason why the inverse ROI data set shows no line feature, while the GC data set shows such a feature? Figure 7 shows that averaged over years, the observing profiles in $cos(LAT\_\theta)$ of the Fermi-LAT depends strongly on the declination ($\delta$) of the ROI being studied. With our current observing strategy of all-sky scanning, the GC collects more time close to the Fermi-LAT z-axis ($cos(LAT\_\theta) \sim 1$) than some other ROIs, and is shown as the black line in the figure. This is because $\delta_{GC} \sim$ inclination of the Fermi-LAT orbit. Also, the observing time distribution for the GC is rather flat in $cos(LAT\_\theta)$. On the other hand, the inverse ROI has a large range of $\delta$ that contributes with some values giving a very difference observing time distribution in $cos(LAT\_\theta)$. Thus, using the inverse ROI as a control sample has additional complexities that require study with Monte Carlo.

.

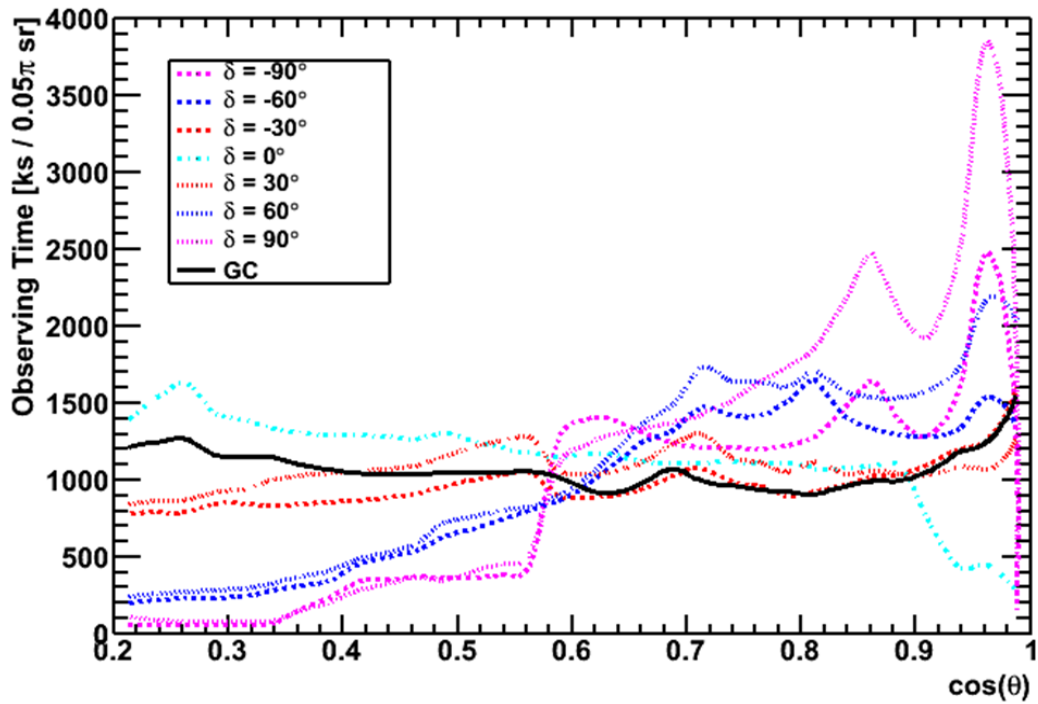

Figure 7: LAT observing profiles in *cos(LAT-θ)* for difference declination (*δ*). Averaged over years, the Fermi-LAT observing profile depends strongly on the *δ* of the ROI. The GC observing profile is shown by the black line. See reference [5].

## 3. SUMMARY AND CONCLUSIONS

### 3.1. Significant feature at 135 GeV

There is a significant feature at 135 GeV in gamma rays from the Earth's limb.

- Feature appears somewhat narrower than the LAT resolution PDF.
- A $LAT\_\theta < 60°$ cut and fitting different energy windows produces different significances for the feature.

### 3.2. Negligible "shine through" of events from non-Earth limb sources

Our selection cuts of $|LRA|>52°$ and $111°<$Earth Zenith angle$<113°$ result in negligible "shine through" of events from non-Earth limb sources, with $\ll 1$ non-limb events in the current sample.

### 3.3. Inverse ROI for P7Rep

- No feature at 135 GeV is evident.
- Given the Earth limb data results, the lack of a 135 GeV feature in the Inverse ROI is currently a bit mysterious.
  - At face value does not support a common systematic that makes a feature at 135 GeV in the gamma-ray data.
- Using the Inverse ROI as a control sample has complexities that need more study with Monte Carlo.

### 3.4. No consistent interpretation of the GC 135 GeV feature

The LAT Collaboration does not have a consistent interpretation of the GC 135 GeV feature originating from a systematic error at this time.

### 3.5. More data and studies are needed to clarify the current situation.

- More detailed analysis with Monte Carlo.
- More Earth limb data.
  - There is now a program of weekly Earth limb following observations for a couple of obits.
  - Target of opportunity (TOO) stares will usually have associated Earth limb following observations when the target source is occulted by the Earth.
- The LAT Collaboration is working on a Pass 8 event reconstruction that is projected to increase the LAT acceptance by 25% above 1 GeV [9].
- The current LAT observing plan will spend more time to collect all-sky data.

**Acknowledgements**

The authors thank members of the Fermi-LAT Collaboration for their work that has been used to produce the study described in this paper.

The Fermi LAT Collaboration acknowledges generous ongoing support from a number of agencies and institutes that have supported both the development and the operation of the LAT as well as scientific data analysis. These include the National Aeronautics and Space Administration and the Department of Energy in the United States, the Commissariat a l'Energie Atomique and the Centre National de la Recherche Scientifque / Institut National de Physique Nucleaire et de Physique des Particules in France, the Agenzia Spaziale Italiana and the Istituto Nazionale di Fisica Nucleare in Italy, the Ministry of Education, Culture, Sports, Science and Technology (MEXT), High Energy Accelerator Research Organization (KEK) and Japan Aerospace Exploration Agency (JAXA) in Japan, and the K. A. Wallenberg Foundation, the Swedish Research Council and the Swedish National Space Board in Sweden. Additional support for science analysis during the operations phase is gratefully acknowledged from the Istituto Nazionale di Astro Fisica in Italy and the Centre National d'Etudes Spatiales in France.